\documentstyle[12pt]{article}
\begin{document}
\def\bt{\begin{tabular}}
\def\et{\end{tabular}}
\def\bfr{\begin{flushright}}
\def\mm{\mbox{\boldmath $ }}
\def\efr{\end{flushright}}
\def\bfl{\begin{flushleft}}
\def\efl{\end{flushleft}}
\def\vs{\vspace}
\def\hs{\hspace}
\def\sta{\stackrel}
\def\pb{\parbox}
\def\bc{\begin{center}}
\def\ec{\end{center}}
\def\sp{\setlength{\parindent}{2\ccwd}}
\def\bp{\begin{picture}}
\def\ep{\end{picture}}
\def\uni{\unitlength=1mm}
\def\REF#1{\par\hangindent\parindent\indent\llap{#1\enspace}\ignorespaces}

\noindent
\bc
{\LARGE\bf Variational-Iterative Solution of \\
\vspace{.2cm}
Ground State for Central Potential}

\vs*{5mm}
{\large  W. Q. Zhao (Chao)$^{1,~2}$}

\vs*{11mm}

{\small \it 1. China Center of Advanced Science and Technology (CCAST)}

{\small \it (World Lab.), P.O. Box 8730, Beijing 100080, China}

{\small \it 2. Institute of High Energy Physics, Chinese Academy of Sciences,

P. O. Box 918(4), Beijing 100039, China}

\ec
\vspace{2cm}
\begin{abstract}

The newly developed iterative method based on Green function defined by 
quadratures along a single trajectory is combined with the variational 
method to solve the ground state quantum wave function for central 
potentials. As an example, the method is applied to discuss the ground 
state solution of Yukawa potential, using Hulthen solution as the trial 
function.
\end{abstract}
\vspace{.5cm}
\noindent
PACS{:~~11.10.Ef,~~03.65.Ge}

\newpage

\section*{\bf 1. Introduction}
\setcounter{section}{1}
\setcounter{equation}{0}

Recently an iterative solution of the ground state for the one
dimensional double-well potential is obtained[1] based on the Green
function method developed in ref.[2]. This Green function is
defined along a single trajectory, from which the ground state
wave function in N-dimension can be expressed by quadratures along
the single trajectory. This makes it possible to develop an
iterative method for obtaining the ground state wave function,
starting from a properly
chosen trial function. The convergence of the iterative solution
very much depends on the choice of the trial function[1].
Therefore the key  of this new method is to find the best trial
function as the starting point of the iteration process.

On the other hand the variational method is a well-known
approximate approach to solve Schroedinger equation starting from a
trial function with a set of variational parameters. By minimizing
the expectation value of the Hamiltonian respect to the set of
parameters of the trial function, the variational solution could be
obtained. If the staring trial function is properly chosen
the solution could be a very good
approximation for the equation. However, it is difficult
to further improve the accuracy of this approximate solution.

In this paper the two methods are combined to get a
veriational-iterative solution. The result is obviously more
accurate than the variational one. It has been shown in ref.[1]
that the final iterative solution is convergent  if
the trial function is properly chosen. As an example the
variational-iterative method is applied to solve Yukawa potential,
using the solution of a Hulthen potential as the trial function.
The obtained result up to the second order is compared to those
based on the pure variational procedure, starting from
the same trial function. The solution of the combined method is
improved significantly.

In Section 2, a brief introduction is given about the Green
function method based on the single trajectory quadrature and the
corresponding iterative formula for central potentials. The
variational-iterative solution for Yukawa potential is given in
Section 3. The numerical results and discussions are given in
Section 4.

\section*{\bf 2. Green Function and Its Iterative Solution \\
for a Central Potential}
\setcounter{section}{2}
\setcounter{equation}{0}

\noindent
{\bf 1. Green function}

We discuss a particle with unit mass, moving in a central
potential $V(r)$. The ground state wave function $\Psi(r)$
satisfies
\begin{eqnarray}\label{e2.1}
[-\frac{1}{2}{\bf \nabla}^2+V(r)]\Psi (r)= E \Psi(r).
\end{eqnarray}
The boundary condition for $\Psi(r)$ is set as
\begin{equation}\label{e2.2}
\Psi(0)=1~~~~~~~{\rm and}~~~~~~~~~~\Psi(\infty)=0
\end{equation}
Now a trial function $\Phi(r)$, satisfying the same
boundary condition as $\Psi(r)$, is introduced into the
following Schroedinger equation:
\begin{equation}\label{e2.3}
[-\frac{1}{2}{\bf \nabla}^2+V_0(r)]\Phi (r)= E_0 \Phi(r).
\end{equation}
The trial function should be chosen such that the difference of
the two potentials
\begin{equation}\label{e2.4}
U(r)=V(r)-V_0(r)
\end{equation}
is small. Defining a Green function $D$ satisfying the following
equation[2]
\begin{equation}\label{e2.5}
[-\frac{1}{2}{\bf \nabla}^2+V_0(r)-E_0]D=1,
\end{equation}
then the solution $\Psi(r)$ can be formally expressed as
\begin{equation}\label{e2.6}
\Psi=\Phi-D(U-\Delta)\Psi
\end{equation}
with $\Delta=E-E_0$. Expressing $\Psi=\Phi f$, then $f$
satisfies
\begin{eqnarray}\label{e2.7}
f &=& 1-\overline{D}(U-\Delta)f
\end{eqnarray}
where
\begin{equation}\label{e2.8}
\overline{D}=\Phi^{-1} D \Phi.
\end{equation}

Following the same procedure as in [2], in the spherical
coordinate system with variables $(r,~\theta,~\phi)$ the single
trajectory is chosen as the radial coordinate $r$ and
correspondingly we define
\begin{eqnarray}\label{e2.9}
h_r^2 &=& [({\bf \nabla} r)^2]^{-1}=1,\nonumber\\
h_{\theta}^2 &=&[({\bf \nabla}\theta)^2]^{-1}=r^2,\\
h_{\phi}^2&=&[({\bf \nabla} \phi)^2]^{-1}=r^2\sin^2 \theta,\nonumber\\
h_{\omega}&=&h_{\theta}h_{\phi}=r^2 \sin \theta\nonumber
\end{eqnarray}
with the volume element
\begin{equation}\label{e2.9a}
d\tau=h_{\omega}dr~d\theta~d\phi.
\end{equation}

Introducing a step function
\begin{equation}\label{e2.10}
(r|\Theta|r') \equiv \left\{
\begin{array}{ll}
1, & ~~~~{\sf for}~~~ r>r'>0\\
0, & ~~~~ {\sf for}~~~ 0<r<r'.
\end{array}
\right.
\end{equation}
or
\begin{equation}\label{e2.10a}
(r|\Theta|r') \equiv \left\{
\begin{array}{ll}
0, & ~~~~{\sf for}~~~ r>r'>0\\
-1, & ~~~~ {\sf for}~~~ 0<r<r'.
\end{array}
\right.
\end{equation}
It can be proved[2] that the Green function is
\begin{eqnarray}\label{e2.11}
\overline{D} &\equiv& -2\Theta\Phi^{-2}\frac{h_r}{h_{\omega}}\Theta\Phi^2 h_r h_{\omega}
\nonumber\\
&=& -2\Theta\Phi^{-2}r^{-2}\Theta\Phi^2 r^2.
\end{eqnarray}
According to the boundary condition of the wave functions we have
\begin{equation}\label{e2.12}
f=1~~~~~~~{\rm at}~~r=0,~~~~~~~{\rm and}~~~~~~~~~
f={\rm finite}~~~~~~~{\rm at}~~r\rightarrow \infty.
\end{equation}
Using the definitions of the step function (\ref{e2.10}) and
(\ref{e2.10a}) to express  the first and the second
$\Theta$-function in (\ref{e2.11}), respectively,
$f$ has the following integral form
\begin{eqnarray}\label{e2.13}
f(r) = 1 -2\int\limits_0^r \Phi^{-2}(r') r'^{-2}dr' \int\limits_{r'}^{\infty} \Phi^2(r'')
(U(r'') - \Delta) f(r'')r''^2dr''
\end{eqnarray}
and $\Psi(r)=\Phi(r)f(r)$ is the solution of eq.(\ref{e2.1}). Considering the definitions
in eqs.(\ref{e2.9}) and (\ref{e2.9a}), the
corresponding energy correction $\Delta$ can be expressed as
\begin{equation}\label{e2.14}
\Delta=\frac{\int\limits_0^{\infty}\Phi^2(r)U(r)f(r)r^2dr}
{\int\limits_0^{\infty}\Phi^2(r)f(r)r^2dr}~,
\end{equation}
which can be formally expressed as
\begin{equation}\label{e2.14a}
\Delta = [U f]/[f],
\end{equation}
with $[F]$ defined as
\begin{eqnarray}\label{e2.14b}
[F] = \int\limits_0^{\infty} \Phi^2(r) F(r)r^2 dr
\end{eqnarray}

\noindent
{\bf 2. Iterative solution}

Based on eqs.(\ref{e2.7}) and (\ref{e2.14a}), two iterative sequences
$\{f_n\}$ and $\{\Delta_n\}$ are introduced as
\begin{eqnarray}\label{e2.15}
f_0 &=& 1,~f_1,~f_2,~f_3,~\cdots,~f_n,~\cdots \nonumber\\
\Delta_0 &=& 0, ~\Delta_1,~\Delta_2,~\Delta_3,~\cdots,~\Delta_n,~\cdots
\end{eqnarray}
We require
\begin{eqnarray}\label{e2.16}
f_n &=& 1 - \overline{D}(U-\Delta_n) f_{n-1}\nonumber\\
\Delta_n &=& [U f_{n-1}]/[f_{n-1}].
\end{eqnarray}
In the explicit form,
\begin{eqnarray}\label{e2.18}
f_n(r) = 1 -2\int\limits_0^r \Phi^{-2}(r')r'^{-2} dr' \int\limits_{r'}^{\infty} \Phi^2(r'')
(U(r'') - \Delta_n)f_{n-1}(r'')r''^2 dr''
\end{eqnarray}
with
\begin{eqnarray}\label{e2.19}
\Delta_n = \int\limits_0^{\infty} \Phi^2(r)f_{n-1}(r) U(r)r^2 dr \bigg/
\int\limits_0^{\infty}\Phi^2(r) f_{n-1}(r)r^2 dr.
\end{eqnarray}
For a properly chosen trial function $\Phi(r)$ this iterative
procedure will approach the solution
\begin{eqnarray}\label{e2.20}
f(r)&=&\lim_{n \rightarrow \infty} f_n(r) ~~~{\rm for~all}~~ r \geq
0,\nonumber\\
\Delta&=& \lim_{n \rightarrow \infty} \Delta_n
\end{eqnarray}
and $\Psi(r)=\Phi(r)f(r)$, $E=E_0+\Delta$.

\section*{\bf 3. Variational-Iterative Solution of Ground State \\
for Yukawa Potential}
\setcounter{section}{3}

The veriational solution of Yukawa potential has been widely
discussed in details[3]. In ref.[3] using a Hulthen solution as the trial
function the result based on variational method is compared to the
results from different methods. In our paper starting from the
Hulthen solution we first give the traditional variational
result. Using it as the trial function of the iterative procedure
further improvement could be obtained. The Hamiltonian for
Yukawa potential is
\begin{equation}\label{e3.1}
H=-\frac{1}{2m}{\bf \nabla}^2-g^2\frac{e^{-\alpha r}}{r}.
\end{equation}
To simplify the notation, based on  the following rescaling[3]:
\begin{eqnarray}\label{e3.2}
\frac{\alpha}{mg^2} &\rightarrow& \alpha\nonumber\\
mg^2 r&\rightarrow& r\\
\frac{E}{mg^4} &\rightarrow& E \nonumber
\end{eqnarray}
the Hamiltonian can be expressed as
\begin{eqnarray}\label{e3.3}
H&=&-\frac{1}{2}{\bf \nabla}^2-\frac{e^{-\alpha r}}{r}\nonumber\\
H\Psi(r)&=&E\Psi(r).
\end{eqnarray}
Introducing the trial function
\begin{equation}\label{e3.4}
\Phi_{\lambda}(r)=\frac{1-e^{-\lambda r}}{r}~e^{-r+\frac{1}{2}\lambda r}
\end{equation}
with a variational parameter $\lambda$, which
satisfies the Schroedinger equation with a Hulthen potential:
\begin{eqnarray}\label{e3.5}
H_{\lambda}&=&-\frac{1}{2}{\bf \nabla}^2
-\frac{\lambda e^{-\lambda r}}{1-e^{-\lambda r}}\nonumber\\
H_{\lambda}\Phi_{\lambda}(r)&=&E_{\lambda}\Phi_{\lambda}(r),
\end{eqnarray}
where
\begin{equation}\label{e3.6}
E_{\lambda}=-\frac{1}{2} +\frac{1}{4} \lambda -\frac{1}{8}\lambda^2.
\end{equation}
Based on the variational method the approximate solution of
eq.(\ref{e3.3}) could be obtained by
minimizing the expectation value of $H$ in eq.(\ref{e3.3}):
\begin{eqnarray}\label{e3.7}
\overline{H}&=&\frac{\int\limits_0^{\infty}\Phi^2_{\lambda}(r)
(-\frac{1}{2}{\bf \nabla}^2-\frac{e^{-\alpha r}}{r}) r^2 dr}
{\int\limits_0^{\infty}\Phi^2_{\lambda}(r) r^2 dr}
\nonumber\\
\frac{\partial \overline{H}}{\partial \lambda}&=&0.
\end{eqnarray}
Now using the obtained wave function $\Phi_{\lambda}(r)$ as the
zero-th order solution, we introduce the formal Green
function expression of $\Psi(r)=\Phi_{\lambda}(r)f(r)$ in the
following way: We can define Green function $D$ and $\overline{D}$
satisfying
\begin{eqnarray}\label{e3.8}
(H_{\lambda}-E_{\lambda})D=1\\
\overline{D}=\Phi_{\lambda}^{-1}D\Phi_{\lambda}.
\end{eqnarray}
As in Section 2, introducing the energy difference
$\Delta=E-E_{\lambda}$ and the potential difference of Hamiltonian
(\ref{e3.3}) and (\ref{e3.5})
\begin{eqnarray}\label{e3.9}
U(r)=\frac{\lambda e^{-\lambda r}}{1-e^{-\lambda r}}-\frac{e^{-\alpha r}}{r},
\end{eqnarray}
we obtain expressions for $f$ and $\Delta$ similar to eqs.(\ref{e2.7}) and
(\ref{e2.14a}). The Green function is
\begin{eqnarray}\label{e3.10}
\overline{D} = -2\Theta\Phi_{\lambda}^{-2}r^{-2}\Theta\Phi_{\lambda}^2 r^2.
\end{eqnarray}
Introducing the iterative series
$\{f_n\}$ and $\{\Delta_n\}$ as (\ref{e2.15}) and (\ref{e2.16}),
the successive terms can be expressed as

\newpage

\begin{eqnarray}\label{e3.11}
\Delta_0&=&0,~~~~~~~~~~~~~~~~~~~~~~~~~~~~f_0=1\nonumber\\
\Delta_1&=&\frac{[U]}{[1]},~~~~~~~~~~~~~~~~~~~~~~~
f_1=1-\overline{D}(U-\Delta_1)\nonumber\\
\Delta_2&=&\frac{[Uf_1]}{[f_1]},~~~~~~~~~~~~~~~~~~~
f_2=1-\overline{D}(U-\Delta_2)f_1\nonumber\\
\cdots\nonumber\\
\Delta_n&=&\frac{[Uf_{n-1}]}{[f_{n-1}]},~~~~~~~~~~~~~~~~
f_n=1-\overline{D}(U-\Delta_n)f_{n-1}.
\end{eqnarray}
Here the trial function for the iterative procedure is the
variational wave function
$\Phi_{\lambda}$. In the explicit form, for the first order, we have
\begin{eqnarray}\label{e3.12}
\Delta_1 &=& \int\limits_0^{\infty} \Phi_{\lambda}^2(r) U(r)r^2 dr \bigg/
\int\limits_0^{\infty}\Phi_{\lambda}^2(r)r^2 dr,\\
E_1&=& E_{\lambda}+\Delta_1\nonumber
\end{eqnarray}
and
\begin{eqnarray}\label{e3.13}
f_1(r) = 1 -2\int\limits_0^r \Phi_{\lambda}^{-2}(r')r'^{-2} dr'
 \int\limits_{r'}^{\infty} \Phi_{\lambda}^2(r'')
(U(r'') - \Delta_1)r''^2 dr''~.
\end{eqnarray}
It is interesting to compare this first order result to the
variational energy and the first order perturbation result. From
eq.(\ref{e3.7})
\begin{eqnarray}\label{e3.14}
\overline{H}&=&\frac{\int\limits_0^{\infty}\Phi^2_{\lambda}(r)
(H_{\lambda}+U)) r^2 dr}
{\int\limits_0^{\infty}\Phi^2_{\lambda}(r) r^2 dr}
\nonumber\\
&=& E_{\lambda}+\Delta_1,
\end{eqnarray}
which is the same as the first order result of the iterative
method. If applying the perturbation method, the first order
perturbation also gives the same energy. However, to further improve the
above result the perturbation method would meet problem with the
unknown excited unperturbed states which are necessary for the
calculation of the cross-matrix elements in the second order calculation.
For the iterative method, in principle, there is no problem for
the higher order calculation. In general, for the n-th order, we
have
\begin{eqnarray}\label{e3.15}
\Delta_n &=& \int\limits_0^{\infty} \Phi_{\lambda}^2(r)f_{n-1}(r) U(r)r^2 dr \bigg/
\int\limits_0^{\infty}\Phi_{\lambda}^2(r) f_{n-1}(r)r^2 dr.\\
E_n&=&E_{\lambda}+\Delta_n,\nonumber
\end{eqnarray}
and
\begin{eqnarray}\label{e3.16}
f_n(r) &=& 1 -2\int\limits_0^r \Phi^{-2}(r')r'^{-2} dr' \int\limits_{r'}^{\infty} \Phi^2(r'')
(U(r'') - \Delta_n)f_{n-1}(r'')r''^2 dr''.
\end{eqnarray}
In the next section the numerical results of $E_1$ and $E_2$ are
given and compared to the result from the variational method for
different $\alpha$ of the Yukawa potential in eq.(\ref{e3.3}).

\section*{\bf 4. Numerical Results and Discussions}
\setcounter{section}{4}

For Yukawa potential with different parameter $\alpha$,
the variational parameter $\lambda$ of a Hulthen potential can be
obtained by minimizing the expectation value of Hamiltonian
eq.(\ref{e3.3}) v.s. $\lambda$. The values of $\lambda$
corresponding to different $\alpha$ are listed in the first two
columns of Table~1. For Hulthen potential the upper limit of
$\lambda$ to have a bound state is $\lambda<2$. Considering this
limitation the highest value of $\alpha$ is taken to be
$\alpha=1.15$ which corresponds to $\lambda=1.9473$. For different
$\{\alpha,~\lambda\}$ pair the obtained zero-th order energy
$E_{\lambda}$, the 1st order energy correction $\Delta_1$ and the
corresponding binding energy of the ground state
$E_1=E_{\lambda}+\Delta_1$ in the first order iteration
are listed in columns 3-5 of
Table~1. As mentioned in Section~3 the values of $E_1$ for
different $\{\alpha,~\lambda\}$ are exactly the same as the variational
result in ref.[3]. In columns 6-7 of Talbe~1 are given the values
of the energy correction $\Delta_2$ and the obtained binding
energy of the ground state $E_2=E_{\lambda}+\Delta_2$ after the second order
iteration. For larger $\alpha$, the correction in the second order
iteration is visible. Comparing to the numerical exact solution in
the column 8 taken from ref.[4] the values after the second order
iteration are obviously improved. In principle, the iteration
could go on to get more accurate result if the
integrations in the iteration procedure could be performed
properly.

It is possible to solve some excited states based on this method
if the proper variational solution could be found.
For example, the variational solution of the excited $S$-states
for Hulthen potential is known[3], the corresponding solution
for Yukawa potential could be obtained by the iteration procedure.
This method could also be extended to multi-dimensional cases as
long as one could choose the proper variational solution and find
the single trajectory to define the Green function and perform the
necessary quadratures[1,2].

\section*{\bf Acknowledgment}

The author is grateful to Professor T. D. Lee for his continuous
and substantial instructions and advice. This work is partly
supported by National Natural Science Foundation of China
(NNSFC, No. 19947001).

\newpage

{\small

\begin{tabular}{|c|c|c|c|c|c|c|c|}
  \hline
  $\alpha$ & $\lambda$ & $-E_{\lambda}$ & $-\Delta_1$ & $-E_1$ & $-\Delta_2$ & $-E_2$ \\
  \hline
  0.1 & 0.2296 & 0.391790 & 0.015628 & 0.407058 & 0.015268 & 0.407058 & 0.4071\\
  \hline
  0.2 & 0.4358 & 0.305840 & 0.020968 & 0.326808 & 0.020968 & 0.326809 & 0.3268\\
  \hline
  0.25 & 0.5327 & 0.269121 & 0.021796 & 0.290918 & 0.021798 & 0.290920 & 0.2909\\
  \hline
  0.3 & 0.6263 & 0.235881 & 0.021753 & 0.257634 & 0.021757 & 0.257639 &\\
  \hline
  0.4 & 0.8052 & 0.178443 & 0.019918 & 0.198362 & 0.019933 & 0.198376 &\\
  \hline
  0.5 & 0.9750 & 0.131328 & 0.016756 & 0.148084 & 0.016789 & 0.148117 & 0.1481\\
  \hline
  0.6 & 1.1377 & 0.0929452 & 0.0131321 & 0.106077 & 0.0131915 & 0.106137 &\\
  \hline
  0.7 & 1.2944 & 0.0622339 & 0.0095107 & 0.0717446 & 0.0096002 & 0.0718341 &\\
  \hline
  0.8 & 1.4461 & 0.0383507 & 0.0062364 & 0.0445871 & 0.0063520 & 0.0447027 &\\
  \hline
  0.9 & 1.5936 & 0.0206451 & 0.00353503 & 0.0241802 & 0.0036677 & 0.0243128 &\\
  \hline
  1.0 & 1.7374 & 0.00861985 & 0.00153843 & 0.0101583 & 0.00165138 & 0.0102712 & 0.01029\\
  \hline
  1.05 & 1.8081 & 0.00460320 & 0.00083757 & 0.00544078 & 0.00093748 & 0.00554068 &\\
  \hline
  1.1 & 1.8780 & 0.00186050 & 0.00034306 & 0.00220356 & 0.00041866 & 0.00227916 &\\
  \hline
  1.15 & 1.9473 & 0.000347161 & 0.000065800 & 0.000412961 & 0.000068335 & 0.000415496 &\\
  \hline
\end{tabular}
}

\bc
{\bf Reference}
\ec

1. R. Friedberg, T. D. Lee and W. Q. Zhao, Ann. Phys. 288(2001)52

2. R. Friedberg, T. D. Lee, W. Q. Zhao and A. Cimenser, Ann. Phys. 294(2001)67

3. C. S. Lam and Y. P. Varshni, Phys. Rev. A4(1971)1875

~~~~~and the papers cited therein.

4. J. Rogers, H. C. Graboske, Jr. and D. J. Harwood, Phys. Rev.
A1(1970)1577

\end{document}